\documentclass[fleqn,twoside]{article}
\usepackage{espcrc2}
\usepackage{graphicx}

\newcommand{\epem}   {\ensuremath{\mathrm{e^+e^-}}}
\newcommand{\qqbar}  {\ensuremath{\mathrm{q\overline{q}}}}
\newcommand{\bbbar}  {\ensuremath{\mathrm{b\overline{b}}}}
\newcommand{\mz}     {\ensuremath{M_{\mathrm{Z}}}}
\newcommand{\as}     {\ensuremath{\alpha_{\mathrm{S}}}}
\newcommand{\asmz}   {\ensuremath{\as(\mz)}}
\newcommand{\asnp}   {\ensuremath{\alpha_{\mathrm{S}}^{\mathrm{NP}}}}
\newcommand{\asq}    {\ensuremath{\as(Q)}}
\newcommand{\oa}     {\ensuremath{\mathcal{O}(\as)}}
\newcommand{\oaa}    {\ensuremath{\mathcal{O}(\alpha_{\mathrm{S}}^2)}}
\newcommand{\oan}    {\ensuremath{\mathcal{O}(\alpha_{\mathrm{S}}^n)}}
\newcommand{\lmqcd}  {\ensuremath{\Lambda_{\mathrm{QCD}}}}
\newcommand{\roots}  {\ensuremath{\sqrt{s}}}
\newcommand{\kperp}  {\ensuremath{k_{\mathrm{t}}}}
\newcommand{\thr}    {\ensuremath{1-T}}
\newcommand{\mh}     {\ensuremath{M_{\mathrm{H}}}}
\newcommand{\mhsq}   {\ensuremath{M_{\mathrm{H}}^2}}
\newcommand{\cp}     {\ensuremath{C}}
\newcommand{\bt}     {\ensuremath{B_{\mathrm{T}}}}
\newcommand{\bw}     {\ensuremath{B_{\mathrm{W}}}}
\newcommand{\fpt}    {\ensuremath{F_{\mathrm{PT}}}}
\newcommand{\cy}     {\ensuremath{c_{\mathrm{y}}}}
\newcommand{\momone}[1] {\mbox{\ensuremath{\langle#1\rangle}}}
\newcommand{\mui}    {\ensuremath{\mu_{\mathrm{I}}}}
\newcommand{\anull}  {\ensuremath{\alpha_0}}

\newcommand{\anulltwo} {\ensuremath{\anull(2~\mathrm{GeV})}}
\newcommand{\cf}     {\ensuremath{C_{\mathrm{F}}}}
\newcommand{\ca}     {\ensuremath{C_{\mathrm{A}}}}
\newcommand{\tf}     {\ensuremath{T_{\mathrm{F}}}}
\newcommand{\nf}     {\ensuremath{n_{\mathrm{f}}}}

\title{ Tests of hard and soft QCD with \epem\ Annihilation Data }

\author{S.  Kluth\address[MPI]{Max-Planck-Institut f\"ur Physik, \\ 
        F\"ohringer Ring 6, 80805 Munich, Germany}}

\begin{document}

\begin{abstract}
Experimental tests of QCD predictions for event shape distributions
combining contributions from hard and soft processes are
discussed.  The hard processes are predicted by perturbative QCD
calculations.  The soft processes cannot be calculated directly using
perturbative QCD, they are treated by a power correction model based
on the analysis of infrared renormalons.  Furthermore, an analysis of
the gauge structure of QCD is presented using fits of the colour
factors within the same combined QCD predictions. 
\vspace{1pc}
\end{abstract}

\maketitle

\section{ INTRODUCTION }

Hadron production in \epem\ annihilation is a well-suited process for
the study of the interplay between hard and soft QCD.  The lack of
interference with the initial state improves the understanding of the
hadronic final state in terms of QCD.  The availability of data sets at
many energy points with centre of mass (cms) energies \roots\ between
14 and 209 GeV allows to separate the hard and soft contributions. 

The hard QCD processes, i.e. the radiation of gluons from the quarks
produced at the electroweak vertex and subsequent higher order
processes are well understood in perturbative QCD
(pQCD)~\cite{ellis96}.  The production of multi-jet events is clear
evidence for the existence of gluons and is successfully described by
pQCD.

The soft contributions to hadron production in \epem\ annihilation are
connected with the transition from the partons (quarks and gluons)
generated by hard processes to the hadrons observed in the
experiment.  This transition is referred to as hadronisation and takes
place at energy scales of approximately a light hadron mass,
i.e. ${\cal O}(100)$~MeV, where pQCD calculations are not reliable due
to the rapid growth of the strong coupling \as.  The soft contributions
are known to scale like an inverse power of the cms energy; for most
observables the scaling is $1/\roots$.

\section{ POWER CORRECTIONS AND INFRARED RENORMALONS }

At energy scales $Q=\lmqcd$ perturbative evolution of \asq\ breaks
down completely due to the Landau pole at \lmqcd.  It is therefore
impossible to attempt pQCD calculations of soft processes without
further assumptions. 

\begin{figure}[!htb]
\begin{center}
\includegraphics[width=0.75\columnwidth]{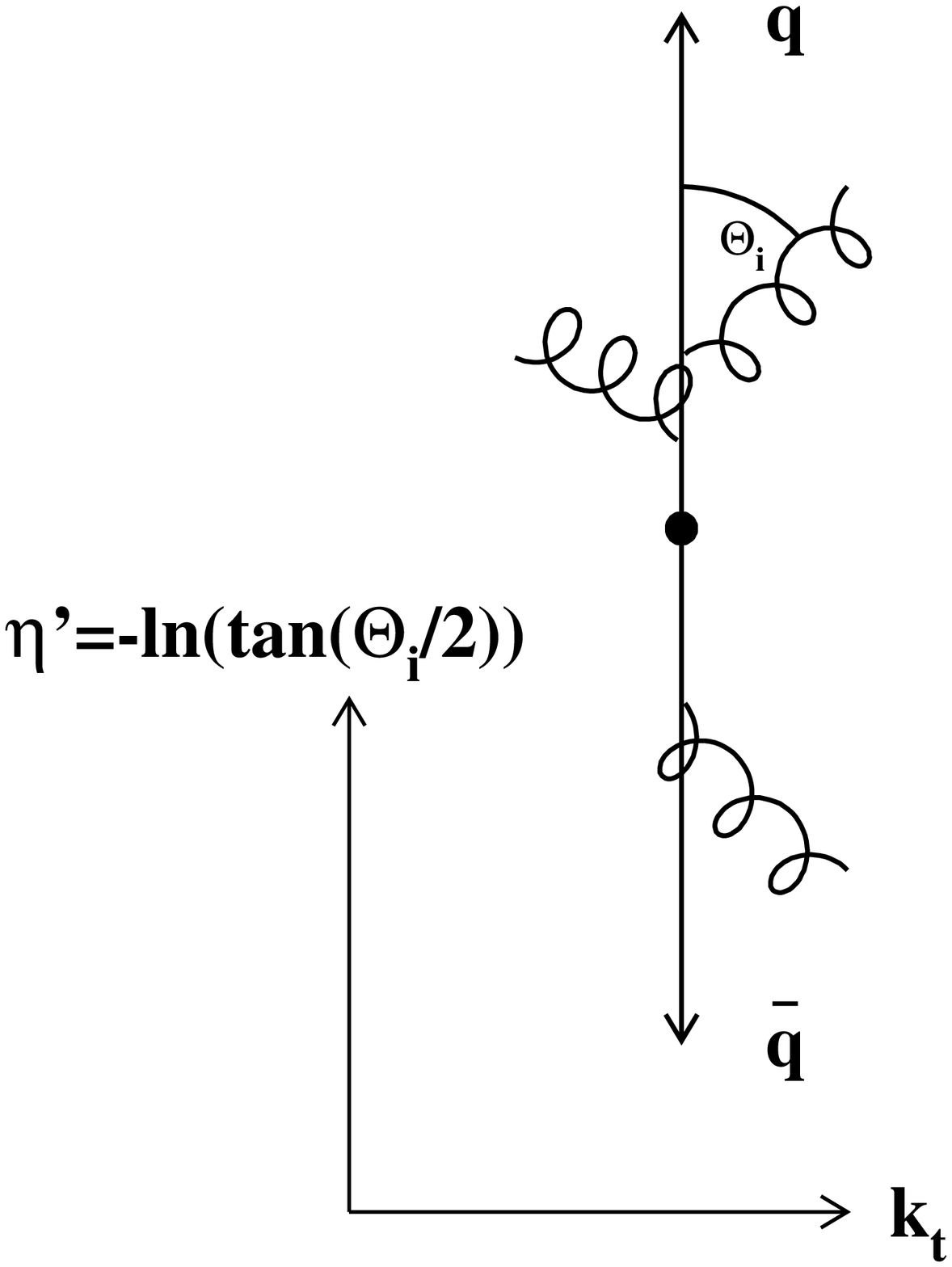}
\caption[ bla ]{ Sketch of a \qqbar\ system in $\eta'-\kperp$ space. }
\label{fig_tube}
\end{center}
\end{figure}

In order to clarify which assumptions are needed a phenomenological
model of hadronisation is considered.  The longitudinal phase space
model or {\em tube model} goes back to Feynman~\cite{salam01a}.  One
considers a \qqbar\ system produced e.g.  in \epem\ annihilation in
the cms system.  The two primary partons move apart with velocity
$v/c\simeq 1$.  In such a situation the production of soft gluons will
be approximately independent of their (pseudo-) rapidity
$\eta'=-\log(\tan(\Theta_i/2))$, see figure~\ref{fig_tube}. 

The change to e.g.  the observable Thrust \thr\ due to
the production of a soft gluon at angle $\Theta_i$ with transverse
momentum $k_{\mathrm{t},i}$ is
$\Delta(1-T)_i\simeq k_{\mathrm{t},i}/Qe^{-|\eta'_i|}$.  
This observation is generalised to
write the soft or non-perturbative contribution to the mean value of
the \thr\ distribution as follows: 
\begin{equation}
  \momone{\thr}_{\mathrm{NP}} = \int\frac{\kperp}{Q}\Phi(\kperp)
                                         \frac{\mathrm{d}\kperp}{\kperp}\cdot 
                                    \int e^{-|\eta'|}d\eta'
\label{equ_tnp}
\end{equation}
The function $\Phi(\kperp)\sim\as(\kperp)$ is the distribution of soft
particles in \kperp.  The first integral in equation~(\ref{equ_tnp}) is
summarised as $\anull/Q$ independent of the observable and the second
as a constant $c_{\thr}$ dependent on the observable.  The quantity
$\anull\sim\int\as(\kperp)\mathrm{d}\kperp$ can only exist when
$\as(\kperp)$ is identified with a non-perturbative strong coupling
$\asnp(\kperp)$ which is assumed to be finite at low \kperp\ around
and below the Landau pole.

A more formal approach to the origin of soft contributions is the
study of infrared renormalons, i.e.  the divergence of asymptotic pQCD
predictions due to the integration of low momenta in quark loops in
gluon lines~\cite{beneke00a}.  The infrared renormalon divergence of
the \oan\ term in a pQCD prediction is factorial in $n$:
$r_n\alpha_S^{n+1}\sim(2\beta_0/p)^n n!\alpha_S^{n+1}$.  With
Sterlings formula to replace $n!$ one finds
$r_n\alpha_S^{n+1}\sim(2\beta_0\as/p)^n n^ne^{-n}\as$.  The
convergence of the series is optimal for $n=p/(2\beta_0\as)$.  Based
on this relation one finds
\begin{equation}
  r_n\alpha_S^{n+1}\sim\left(\frac{\lmqcd}{Q}\right)^p
\label{equ_renorm}
\end{equation}
where the first order relation between \as\ and \lmqcd\ has been
used.  The result shows that infrared renormalon contributions to pQCD
predictions scale like $Q^{-p}$ similar to the soft contribution
studied in the tube model, see equation~(\ref{equ_tnp}). 

The power correction model of Dokshitzer, Marchesini and Webber (DMW)
extracts the structure of power correction terms from analysis of
infrared renormalon contributions~\cite{dokshitzer95a}.
The model assumes that a non-perturbative strong coupling exists
around and below the Landau pole and that the quantity
$\anull(\mui)=1/\mui\int_0^{\mui}\asnp(\kperp)\mathrm{d}\kperp$ can be
defined.  The value of \mui\ is chosen to be safely within the
perturbative region, usually $\mui=2$~GeV.

The main result for the effects of power corrections on distributions
$F(y)$ of the event shape observables \thr, \mh\ and \cp\ is that 
the perturbative prediction $\fpt(y)$ is
shifted~\cite{dokshitzer97a,dokshitzer98b,dokshitzer99a}: 
\begin{equation}
  F(y)= \fpt(y-\cy P)
\label{equ_npshift}
\end{equation}
where \cy\ is an observable dependent constant and $P\sim
M\mui/Q(\anull(\mui)-\as)$ is universal, i.e. independent of the
observable~\cite{dokshitzer98b}.  The factor $P$ contains the $1/Q$
scaling and the so-called Milan-factor $M$ which takes two-loop
effects into account.  The non-perturbative parameter \anull\ is
explicitly matched with the perturbative strong coupling \as.  For the
event shape observables \bt\ and \bw\ the predictions are more
involved and the shape of the pQCD prediction is modified in addition
to the shift~\cite{dokshitzer99a}.  For mean values of \thr, \mh\ and
\cp\ the prediction is:
\begin{equation}
  \momone{y}= \momone{y}_{\mathrm{PT}}+\cy P
\label{equ_pcmean}
\end{equation}
For \momone{\bt} and \momone{\bw} the predictions are also more
involved.

\section{ EXPERIMENTAL TESTS }
\label{sec_exp}

Distributions and mean values of event shape observables measured in
\epem\ annihilation at many cms energy points between 14 and more than
200 GeV have been used recently by several groups to study power
corrections~\cite{powcor,l3lep2data2,delphilep2data3,alephlep2data1}.
This report will present mainly results from~\cite{powcor}.  In all of
these studies pQCD predictions in
\oaa+NLLA~\cite{nllathmh,nllabtbw2,nllacp} are used for event shape
distributions while \oaa~\cite{yellow1qcd} predictions are employed
for mean values.  The pQCD predictions are added together with the
power correction predictions.  The resulting expressions are functions
of the strong coupling \asmz\ and of the non-perturbative parameter
\anull.  The complete predictions are compared with data for event
shape distributions or mean values as published by the experiments.

\begin{figure}[!htb]
\begin{center}
\includegraphics[width=\columnwidth]{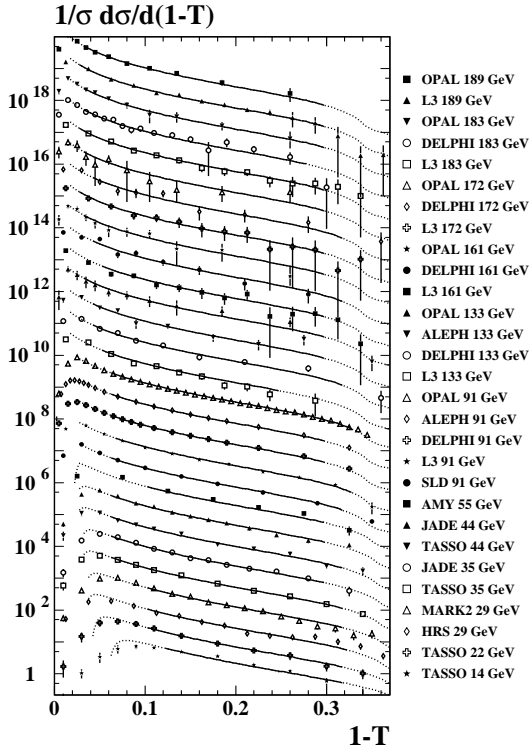}
\caption[ bla ]{ Scaled distributions of \thr. The solid lines show
the result of the fit of the combined pQCD and power correction
prediction.  The dotted lines represent an extrapolation of the
result~\cite{powcor}. }
\label{fig_thdist}
\end{center}
\end{figure}

Figure~\ref{fig_thdist} shows the results fitting the data for \thr\ 
distributions measured at cms energies from 14 to 189~GeV.  The data
from experiments at $\roots<\mz$ are corrected for the effects of
$\epem\rightarrow\bbbar$ events~\cite{powcor}.  The data in
the fitted regions are well reproduced by the theory (solid lines).
The dotted lines present extrapolations outside of the fitted regions
using the fit results for \asmz\ and \anull\ which describe the data
reasonably well.  The agreement between theory and experiment is
similar for the other observables. 

\begin{figure}[!htb]
\begin{center}
\includegraphics[width=0.8\columnwidth]{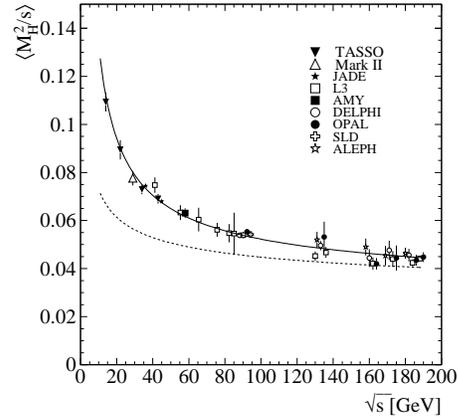}
\caption[ bla ]{ The cms energy dependence of \momone{\thr} is shown.
The solid lines show the result of the fit of combined \oaa\ QCD
calculations with power corrections while the dashed lines indicate
the perturbative contribution~\cite{powcor}. }
\label{fig_mhmean}
\end{center}
\end{figure}

Figure~\ref{fig_mhmean} (solid line) presents the result of a fit to
measurements of \momone{\mhsq} at $\roots=14$ to 189~GeV.  The data are
well described by the fitted theory.  The dotted line in
figure~\ref{fig_mhmean} shows the perturbative part of the prediction;
one observes that the soft (power correction) contributions increase
at low \roots. 

\begin{figure}[!htb]
\begin{center}
\begin{tabular}{c}
\includegraphics[width=0.8\columnwidth]{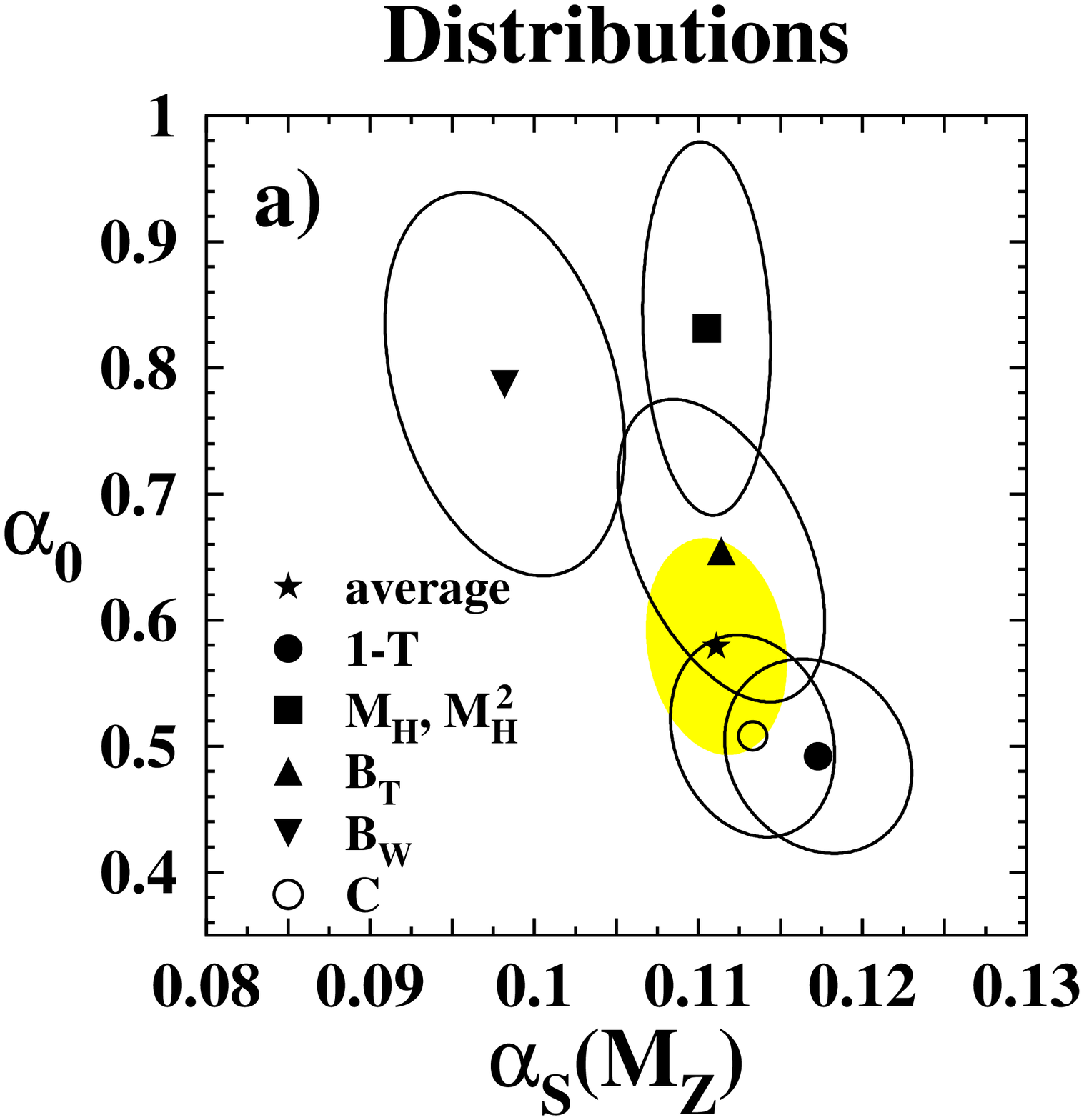} \\
\includegraphics[width=0.8\columnwidth]{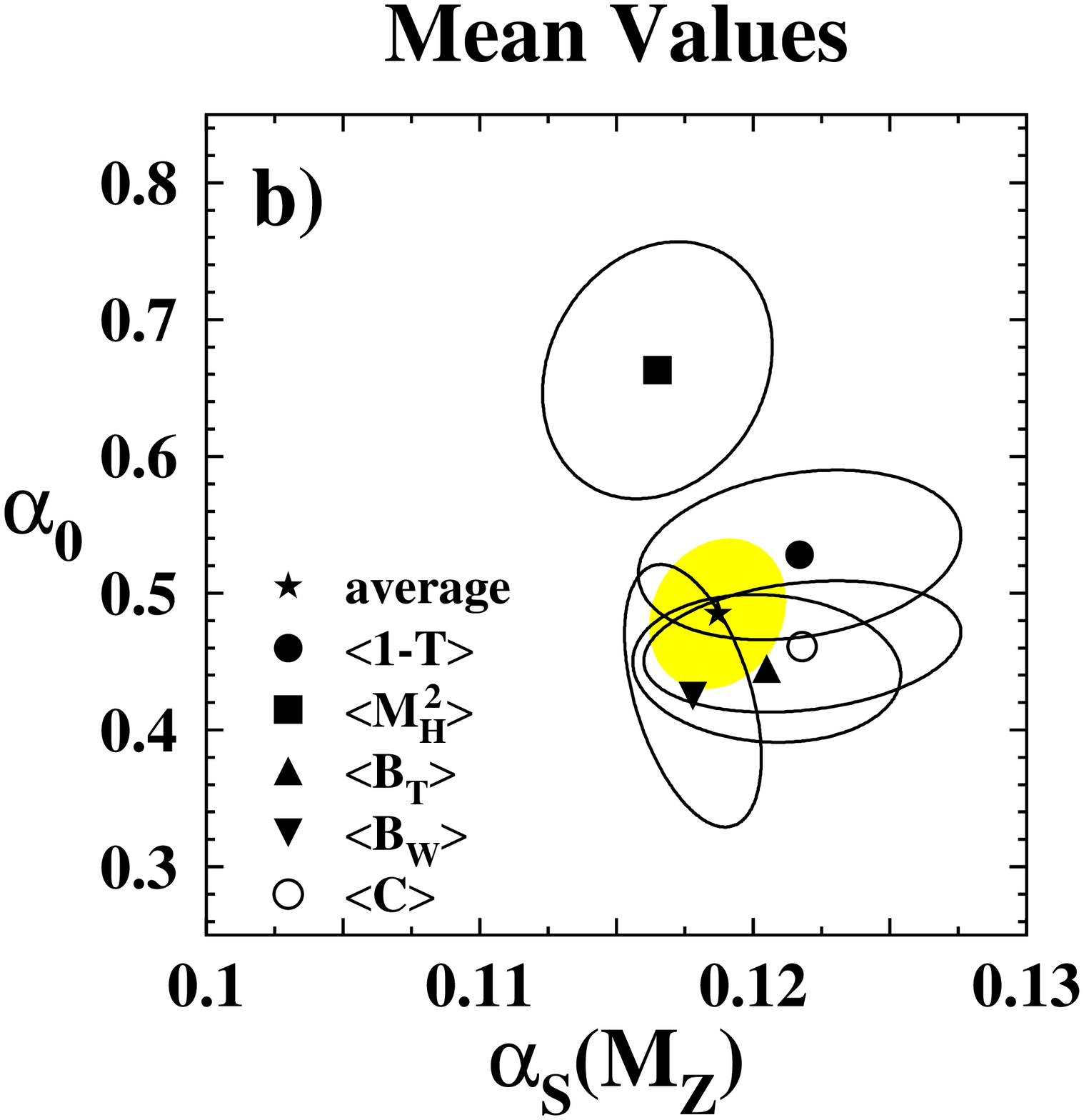}
\end{tabular}
\caption[ bla ]{ Results for \asmz\ and \anulltwo\ for distributions
(top) and mean values (bottom).  The error ellipses show one standard
deviation uncertainties (38\% CL)~\cite{powcor}. }
\label{fig_asa0}
\end{center}
\end{figure}

The combined results from the analysis of distributions are
$\asmz=0.1111^{+0.0048}_{-0.0037}$ and
$\anulltwo=0.579^{+0.100}_{-0.072}$.  The relatively small value of
\asmz\ compared e.g. to the world average
$\asmz=0.1184\pm0.0034$~\cite{bethke00a} is due to small results from
\mh\ and \bw.  The individual and the combined results are shown as
points with one-standard-deviation error ellipses in
figure~\ref{fig_asa0} a). 

From the analysis of mean values the combined results are
$\asmz=0.1187^{+0.0031}_{-0.0020}$ and
$\anulltwo=0.485^{+0.066}_{-0.045}$ in reasonable agreement with the
results from distributions.  Figure~\ref{fig_asa0} b) presents the
individual and combined results from fits to mean values as points
with one-standard-deviation error ellipses.  Combining both analyses
yields the final results: $\asmz=0.1171^{+0.0032}_{-0.0020}$ and
$\anulltwo=0.513^{+0.066}_{-0.045}$ .

\begin{figure}[!htb]
\begin{center}
\includegraphics[width=0.8\columnwidth]{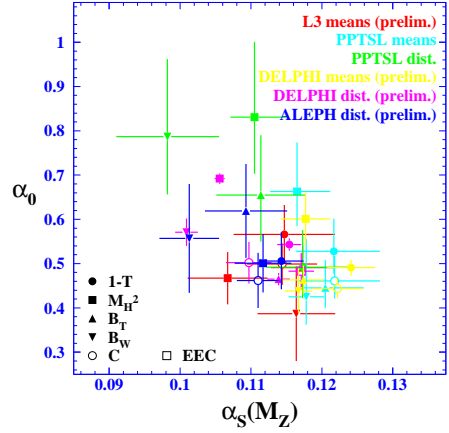}
\caption[ bla ]{ The figure shows a summary of results for \asmz\ and
\anulltwo\ from power correction
studies~\cite{powcor,l3lep2data2,delphilep2data3,alephlep2data1}.  The
results labelled PPTSL are from~\cite{powcor}. } 
\label{fig_allasa0}
\end{center}
\end{figure}

Figure~\ref{fig_allasa0} shows a summary of results for \asmz\ and
\anulltwo\ obtained in comparable analyses by several
groups~\cite{powcor,l3lep2data2,delphilep2data3,alephlep2data1}.  One
observes reasonable agreement between the individual measurements.
The comparable results for \mh\ and \bw\ from~\cite{powcor}
and~\cite{delphilep2data3} differ, because in~\cite{delphilep2data3}
hadron mass effects on power corrections are
considered~\cite{salam01a}.  The non-perturbative parameter \anull\
appears universal within about 20\% as expected~\cite{dokshitzer98b}
while the values for \asmz\ are compatible with the world
average~\cite{bethke00a}.

\section{ QCD COLOUR FACTORS }

The analysis of event shape distributions described above can be
generalised to study the gauge structure of QCD~\cite{colrun}.  The
\oaa+NLLA pQCD predictions can be decomposed into additive terms
which are proportional to products of the QCD colour factors.  This
analysis constitutes an experimental test of the gauge symmetry of QCD
via radiative corrections.  The values of the colour factors
correspond to the relative size of contributions from three
fundamental processes i) gluon radiation from a quark (\cf), ii)
conversion of a gluon in a \qqbar\ pair (\tf\nf) and iii) radiation of
a gluon from a gluon (triple gluon vertex TGV) (\ca).  Higher order
processes like the quartic gluon vertex are not accessible with the
available pQCD calculations.

\begin{figure}[!htb]
\begin{center}
\begin{tabular}{c}
\includegraphics[width=0.8\columnwidth]{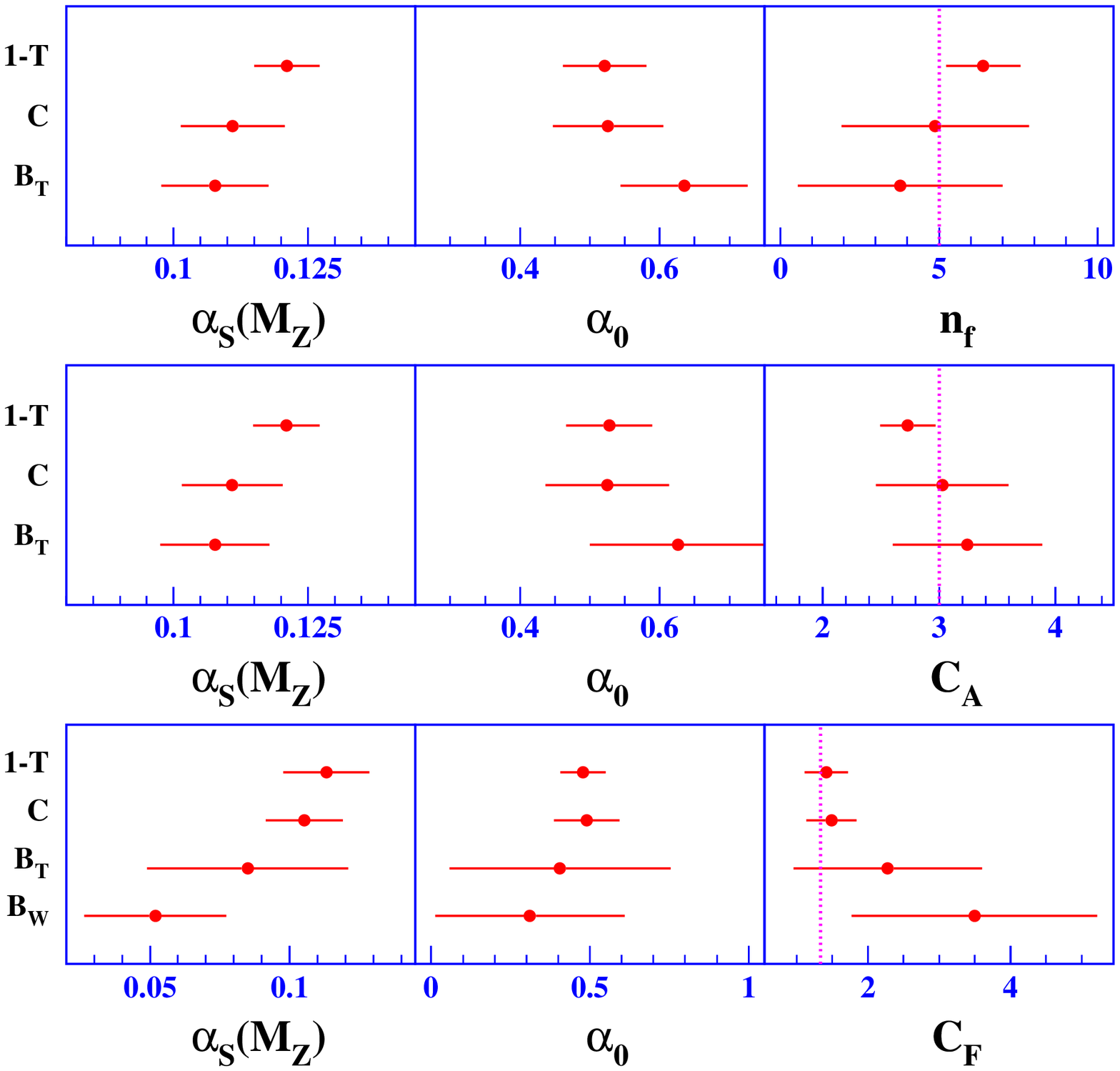} \\
\includegraphics[width=0.8\columnwidth]{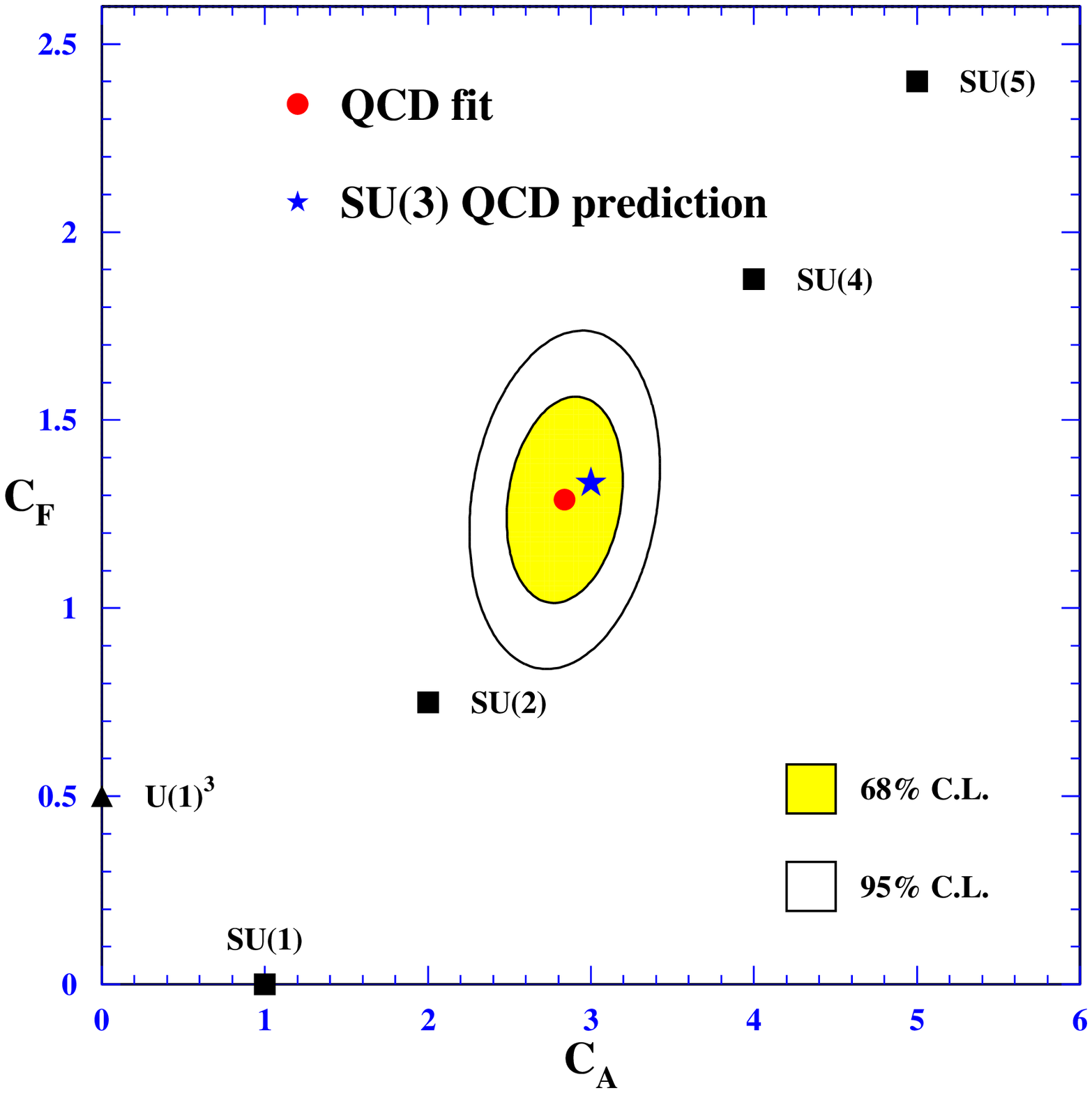}
\end{tabular}
\caption[ bla ]{ The figure at the top shows results for \asmz,
\anulltwo\ and one of the colour factors for individual observables as
indicated.  The error bars show total uncertainties and the dotted
lines show the expectation from SU(3) QCD.  The figure at the bottom
presents combined results for \ca\ and \cf\ from fits to \asmz, \ca\
and \cf\ with \thr\ and \cp.  The square and triangle symbols indicate
expectations for \ca\ and \cf\ for several symmetry
groups~\cite{colrun}. } 
\label{fig_colrun}
\end{center}
\end{figure}

The values of the colour factors are determined in the theory when a
particular gauge symmetry group is chosen under which the Lagrangian
of theory must be locally invariant.  The local gauge symmetry for QCD
is SU(3) with $\cf=4/3$, $\ca=3$ and $\tf=1/2$.  The number of active
quark flavours is $\nf=5$ at the cms energies of the currently
available data.  The perturbative evolution (running) of the strong
coupling is also sensitive to the colour factors since quark loops and
gluon loops have the opposite effect.  In \oa\ the evolution of the
strong coupling between two scales $Q$ and $\mu$ is given by
$\asq=\as(\mu)/(1+2\beta_0\as(\mu)\ln(Q/\mu))$ with
$\beta_0=(11\ca-2\nf)/(12\pi)$.

The power correction calculations are also functions of the QCD colour
factors.  In the analysis the colour factors in the
pQCD predictions as well as in the power corrections are considered as
additional free parameters to be determined from the data.  In this
way the dependence on the hadronisation model is reduced compared to
traditional analyses relying on Monte Carlo models of hadronisation,
because in the Monte Carlo models the colour factors are
fixed~\cite{aleph4jet2,4jetdelphi3,OPALPR330}.  

Figure~\ref{fig_colrun} (top) shows the results of simultaneous fits of
\asmz, \anulltwo\ and one colour factor where the other colour factors
have been kept fixed at their SU(3) values.  The fits are stable with
the observables \thr\ and \cp.  Some fits are not stable with the
observables \bt\ and \bw.  In alternative fits the non-perturbative
parameter \anull\ was kept fixed at a previously measured value and is
varied by its total error as a systematic uncertainty.  The results
and the uncertainties are consistent between the two types of fit.
The variation of \anull\ in the fits avoids a possible bias which
might be present when \anull\ is fixed to a value measured with colour
factors fixed to their standard values.  Combined results for the
colour factors measured individually are $\nf=5.64\pm1.35$,
$\ca=2.88\pm0.27$ and $\cf=1.45\pm0.27$.  The corresponding results
for \asmz\ and \anull\ are consistent with previous measurements. 

Figure~\ref{fig_colrun} (bottom) presents the combined results of
simultaneous fits to \thr\ and \cp\ of \asmz, \ca\ and \cf\ with
\anull\ fixed to previously measured value.  Such fits with \anull\ as
a free parameter turn out to be unstable mainly due to limited
precision of the data.  The results are $\ca=2.84\pm0.24$ and
$\cf=1.29\pm0.18$ with a consistent value of $\asmz=0.119\pm0.010$ and
consistent with the expectation from SU(3).  Several other possible
gauge groups are excluded.

\section{ SUMMARY }

The analyses of power corrections discussed in section~\ref{sec_exp}
show that it is possible to describe event shape distributions and
mean values in \epem\ annihilation data with combined pQCD and power
correction predictions without the need for Monte Carlo based
hadronisation corrections.  Global fits of the combined theory with
only \asmz\ and \anulltwo\ as free parameters to event shape data
measured at $\roots=14$~GeV up to the highest LEP~2 energies lead to a
satisfactory agreement with the data.  The non-perturbative parameter
\anulltwo\ is found to be universal, i.e. independent of the
observable, within the expected theoretical uncertainty of about 20\%. 

A generalisation of the power correction analysis to also vary the QCD
colour factors provides a measurement of the colour factors based on
the QCD radiative corrections in the pQCD and the power correction
predictions.  The results for the colour factors are consistent with
the expectations from SU(3) as the gauge group of QCD while the total
uncertainties are competitive with analyses of angular correlations in
4-jet final states in \epem\ annihilation. 

One of the legacies of LEP, SLC and their predecessors TRISTAN, PETRA
and PEP is the wealth of measurements using hadronic final states in
\epem\ annihilation.  These data made is possible to test and verify
combined predictions for hard (pQCD) and soft (power corrections)
processes.  Based on these tests our understanding of the hard and
soft processes and their interplay has improved significantly. 

The author would like to take the opportunity to thank the organisers
for a stimulating and entertaining meeting.


\end{document}